\documentclass[conference]{IEEEtran}
\usepackage{amsmath}
\usepackage{graphicx}
\usepackage{url}
\usepackage{mcode}
\usepackage{amsmath}
\usepackage{color}
\usepackage{xfrac}
\usepackage{flushend}
\usepackage{subfiles}
\usepackage{float}

\makeatletter
\renewcommand*\env@matrix[1][*\c@MaxMatrixCols c]{%
  \hskip -\arraycolsep
  \let\@ifnextchar\new@ifnextchar
  \array{#1}}
\makeatother

\begin{document}

\title{Vehicle Guidance and Tracking Systems}
\author{
\IEEEauthorblockN{Ryan Baker, John Garvey, 
Mitchell Kraft, Manoj Mathews, 
Kieran O’Connor, Matthew Wolf
\\}
\IEEEauthorblockA{Command and Control 09425\\
05/07/2020\\
Emails: (bakerr5, garveyj7, kraftm4, mathew48, oconnork9, wolfm6)@students.rowan.edu}
}
\maketitle

\IEEEpeerreviewmaketitle

\begin{abstract}
Our application of command and control is the Aegis Combat System. Major components of this system include: missile guidance and missile tracking. To look further into some of the aspects of these systems, an extremely simplified model of the Aegis Combat System will be designed. In this simplified model, a small-scale car will autonomously follow a small-scale remote controlled car. There will be three major components of this system:the controller and the two small-scale cars. Through this model, the team can demonstrate the real world application of certain aspects of C2 such as command, communication, and sensor data fusion. Figure 1 shows a picture of the Aegis Combat System. 
\begin{figure}[H]
\centering
\includegraphics[width=.8\linewidth]{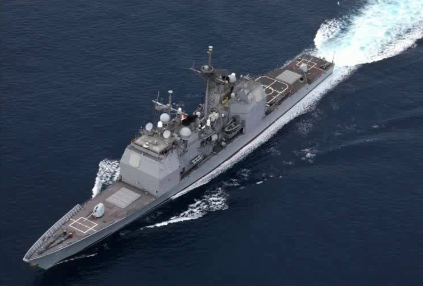}
\caption{Aegis Combat System}
\label{Figure1a1b}
\end{figure}
\end{abstract}

\section{Introduction}
The Aegis Combat System (ACS) was initially developed by the Missile and Surface Radar Division of RCA, and then being sold to various companies until becoming a part of Lockheed Martin in 1995. ACS uses powerful computers and radars to track and guide weapons to destroy enemy targets. The total cost of the ACS system is approximately \$42.7 billion [9]. To put it in perspective when the Navy converted the cruiser Normandy into an Aegis Baseline 9 ship, with the updated displays and blade servers, the cost of the service was about \$188 million and nearly a year offline [10]. The goal of our project is to highlight the real world components of the ACS which are command, communications, and sensor data fusion. The command application controls the strategies that are given to the driver in order to see how different evasion strategies are more or less effective at shaking off the second vehicle. Then we have communication where first vehicle communicates with the controller over Bluetooth where the effects of latency, throughput, and range can be observed.The limitations of Bluetooth may be fatal in certain environments. Lastly we have sensor data fusion where several sensors will be used for the second vehicle to track the first vehicle. How the data is collected and combined will play into how successful the second vehicle is at tracking the first vehicle.

\section{Description of Project}
The ACS system has a variety applications that can be used such as: 
\begin{itemize}
  \item Determining the location of a vehicle at any given period of time
  \item Optimize driver routes to save petrol, gas, or time
  \item Reduce theft and control the vehicle functions
  \item Mobility pattern recognition 
  \item Vehicle navigation
  \item Fleet management
  \item Route tracking
\end{itemize}
 By using these applications in more simplified form, we plan on simulating a pursuit between a small-scale autonomous car tracking a small-scale remote controlled car. The autonomous car should be able to follow the remote control car until the gap is closed where the cars are touching each other. In order to simulate this the remote controlled car with a light emitting device attached to the top of the car while the autonomous vehicle will have a device attached to the top of it that tracks light.
 
 \subsection{Real World Application}
 A form of command guidance used from the 1980s by the U.S. Patriot surface-to-air system was called track-via-missile. In this system a radar unit in the missile tracked the target and transmitted relative bearing and velocity information to the launch site, where control systems computed the optimal trajectory for intercepting the target and sent appropriate commands back to the missile. As you can see in figures 2 and 3 we have two cases: direct flight to intercept and tail chase to intercept. Figure 2 displays direct flight to intercept where the missile is shot via ship to a specific coordinate where it will direct impact the head of the incoming missile disarming it.
\begin{figure}[H]
\centering
\includegraphics[width=.8\linewidth]{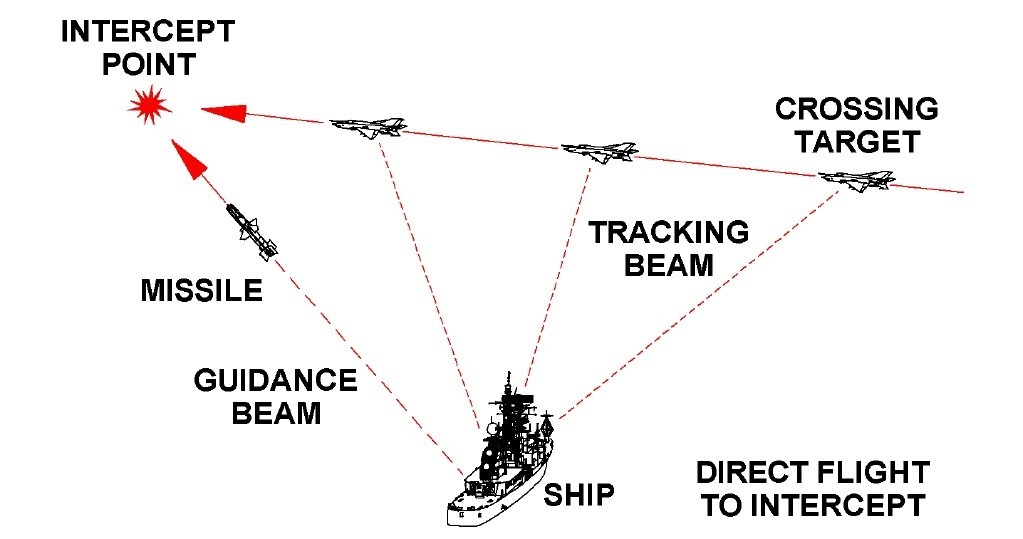}
\caption{Direct Flight to Intercept}
\label{Figure1a1b}
\end{figure}

Figure 3 shows the second case if the specific coordinates cannot be determined or a miscalculation is done, the intercepting missile follows "Tail Chase to Intercept" command. 
\begin{figure}[H]
\centering
\includegraphics[width=.8\linewidth]{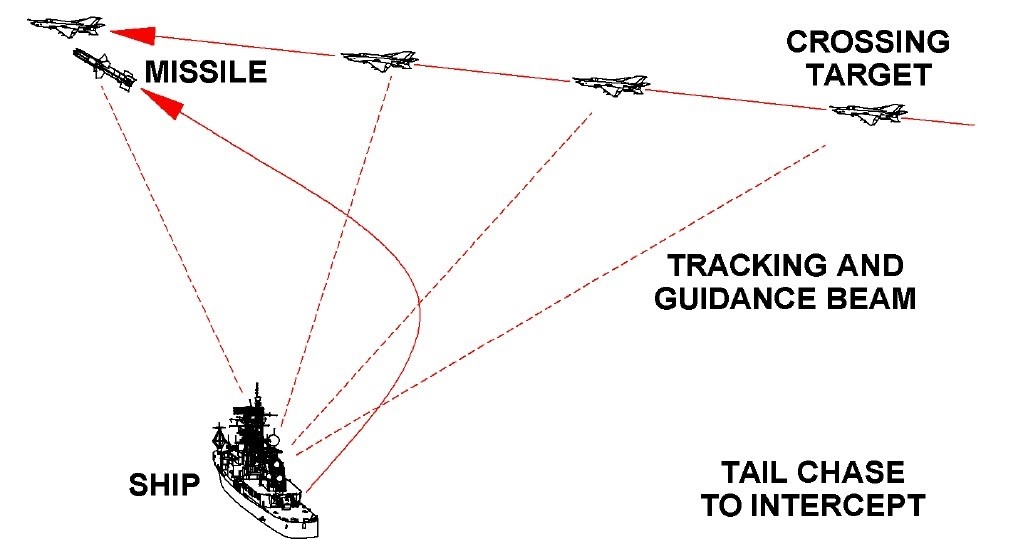}
\caption{Tail Chase to Intercept}
\label{Figure1a1b}
\end{figure}

\section{History of Problem}
% Section on the history of the problem
Inertial guidance systems for ballistic missiles is the earliest form of technology related to guidance and tracking systems. Dr. Fritz Mueller details the history of inertial guidance in his report for the Defense Documentation Center for Scientific and Technical Information, which provides information on the history of such systems all the way back to the early 1930s.

Inertial guidance is a self-contained system in which the payload possesses the ability to guide and control itself without assistance from a ground controller. The guidance of this payload is able to sense and correct itself from outside forces such as side winds. The earliest form of this technology uses gyroscopic principles that were derived from Newton's laws of motion. The first ever technological implementation of this principle was conducted by the Germans in 1934.

The first rocket developed by the Germans which utilized such guidance technology is the A-2 rocket which contained a rudimentary gyro system. This system was placed between the oxygen and fuel containers on the rocket directly into the center of the entire system. 

This gyro wheel system was allowed to coast freely during flight, allowing the rocket to correct its altitude by a simple brute force technique. Only two of these rockets were launched in 1934, as it was discovered that this system contained numerous shortcomings. The large weight and mounting system used caused the A-2 to have inter coupling problems between the system and payload.

The Germans continued to iterate on this design, and by the end of the 1930s reached the A-5 Rocket. This rocket used a much simpler guidance system, which contained a three-gyro and three-axis stabilized platform which controlled not only altitude but also tilt of the rocket. This rocket also contained a unique control system that was able to take in angular deviation signals that were sensed by the gyros in order to better control the rocket during flight. A detailed cross-section of this rocket can be seen in the figure below. 

\begin{figure}[H]
    \centering
    \includegraphics[scale=0.5]{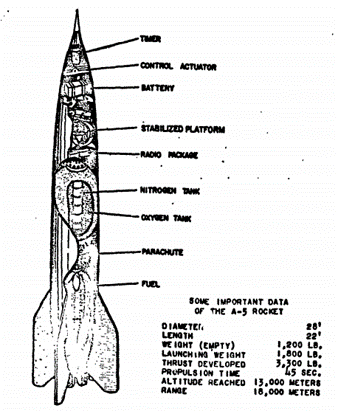}
    \caption{Cross Section of the German A-5 Rocket}
    \label{Figure3b}
\end{figure}

The first German A-5 missiles would be launched in late 1939 and ultimately served as a precursor to the V-2 rocket. This rocket utilized the LEV-3 guidance system which contained two free gyros, control potentiometers, pendulums, servo motors, and other mounting and adjustment devices. The two gyros in this system each had distinct jobs, one controlling the yaw and roll deviations and the other controlling the pitch and tilt deviations of the rocket. This was the most dependable guidance system developed by the Germans during this time \cite{mueller}. 

Moving forward to a modern application of guidance and tracking systems with the launch of AEGIS in 1969 with the Lockheed Martin legacy company RCA \cite{Aegislock}. This system is able to combat short to intermediate range ballistic missile threats with Standard Missile-3, or SM-3; as well as, short range ballistic missile threats with the terminal phase Standard Missile-2, or SM-2 \cite{BMDmil}. Aegis Ballistic Missile Defense, or BMD, ships detect and track missiles of all ranges and report this data back to the missile defense system. Unlike traditional radar systems, AEGIS locks on to multiple threats or aircraft simultaneously, tracking their movements and assessing every entity based on a threat level. This allows the crew members or other human elements to asses and engage threats with proper countermeasures. In 1974 this system was tested aboard the USS Norton Sound and was able to successfully intercept aerial targets along the Pacific Missile Test Range \cite{Aegislock}.

Missiles and ballistics represent incredibly tracking systems and continue to be improved upon as time goes on; however, the most recent and impressive form of vehicle guidance is none other than the Tesla Autopilot system. This autopilot system is present in every current Tesla model electric vehicle through advanced hardware and constant software updates. This system contains eight individual cameras that provide 360 degrees of visibility around the car up to 250 meters away. Twelve additional ultrasonic sensors allow for both hard and soft object detection at nearly 500 meters. Finally, each vehicle is equipped with a forward facing radar with enhanced processing that is able to provide redundant data to the system, allow the computer vision to see through rain, fog, and even dust. This system is fully capable of navigating passengers to their destination with optimized routes, steering through narrow, tight, or even complex roads, and also possesses the ability to be summoned to the driver from parking spaces. This system is truly at the forefront of navigational guidance \cite{tesla}.

\begin{figure}[H]
    \centering
    \includegraphics[scale=0.425]{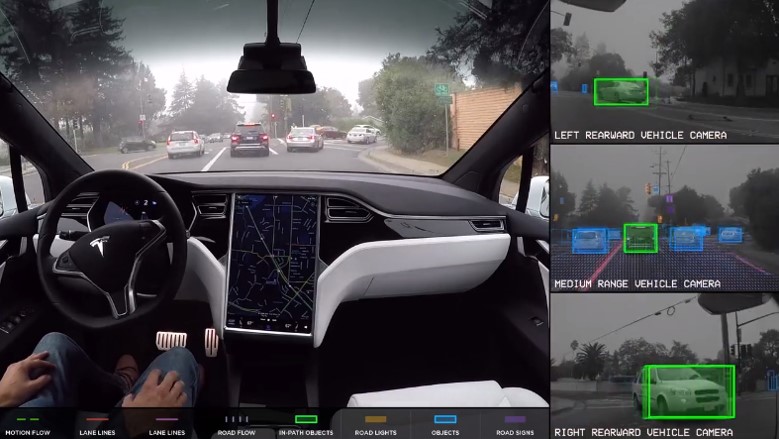}
    \caption{Tesla Autopilot and Tesla Vision in Action}
    \label{Figure3c}
\end{figure}

\section{Core C2 Functions}
% Section on the Core C2 Functions that you have investigated (of the seven)
The seven C2 core functions include main ideas of a revolving around a system. The seven include intent, roles, rules, monitoring, inspiring, training, and provisioning. Starting with intent in general when discussing a guiding system the goal is to get the desired object to a designated target. In this case the simulation is the follower and approaches the lead car. The roles in the system include the pursuer, the target, and the guider. The purser approaches the target and the guider directs the pursuer to the target. Rules or constraints could include many things for example the developed simulation if it loses sight of the target the pursuer continues in the last know direction of the target. Monitoring the updated progress of the system makes sure it is functioning and is still closing in on the desired target. Inspiring or trusting the system could go hand in hand with training since the more developed the system is the more the system can be relied on to do its job. This can also relate to the last function provisioning resource to test and continuously improve a guiding system such as the Aegis Combat System making sure it is calibrated and the best it can operate.

\section{Model Design}
% Section on the model design (Physical-Information-Cognitive-Social)
%  – Consider the Shannon information issues (entropy, etc.)
%  – Show Ashby control considerations
%  – Consider Conceptual Spaces and how these map
Model design consists of four main domains physical, informational, cognitive, and social. Physical collects the base location of the pursuer and target of the system. The informational domain continuously stores this data for future use. The cognitive domain uses the objective and information to gain an understanding of the situation then in the case of the Aegis Combat System controllers determines the proceeding actions to take in order for the pursuer to collide with the target. Another example of the cognitive domain is shown in the developed simulation. The cognitive domain is comprised of the driver controlling the leader remote car determining where the car will go. The cognitive domain then uses the remote to transmit directions to the car this is the social domain.

\subsection{Ashby Control Considerations}
\begin{figure}[H]
    \centering
    \includegraphics[width=.7\linewidth]{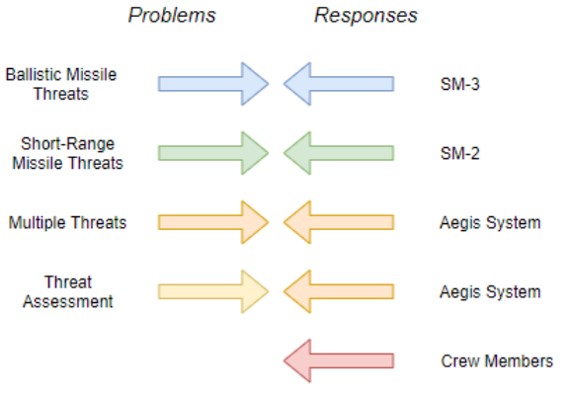}
    \caption{Ashby Control Considerations for AEGIS}
    \label{Figure5a}
\end{figure}

The AEGIS weapon system is a robust system with numerous amounts of redundancy included. Ashby's Law of Requisite Variety is used to determine when a system is stable, specifically the number of states that the control mechanism can contain must be greater than those being controller. A top level look at the control considerations can be seen in Figure \ref{Figure5a} above. The AEGIS system has been determined to possess more response than problems for the system, as each type of threat is met with a specific type of missile and multiple amounts of threats are addressed by the software side of the system. It has been determined that the added crew members or any individual who also possesses input to the system adds redundancy to the entire system. These individuals are able to make decisions based on parameters outside of the system, adding to its robustness. These top level ideas alone prove why the AEGIS Missile Ballistic Defense System satisfies the Ashby control considerations. 

\section{SoS Considerations}
% Section for the systems or SoS considerations

The guidance and tracking system as a whole is an incredibly complex system.  The system itself can be described as a fusion of inter-operable systems with a common goal.  This makes the system a system of systems (SoS).  

\subsection{System Breakdown}
\begin{figure}[H]
    \centering
    \includegraphics[scale=0.25]{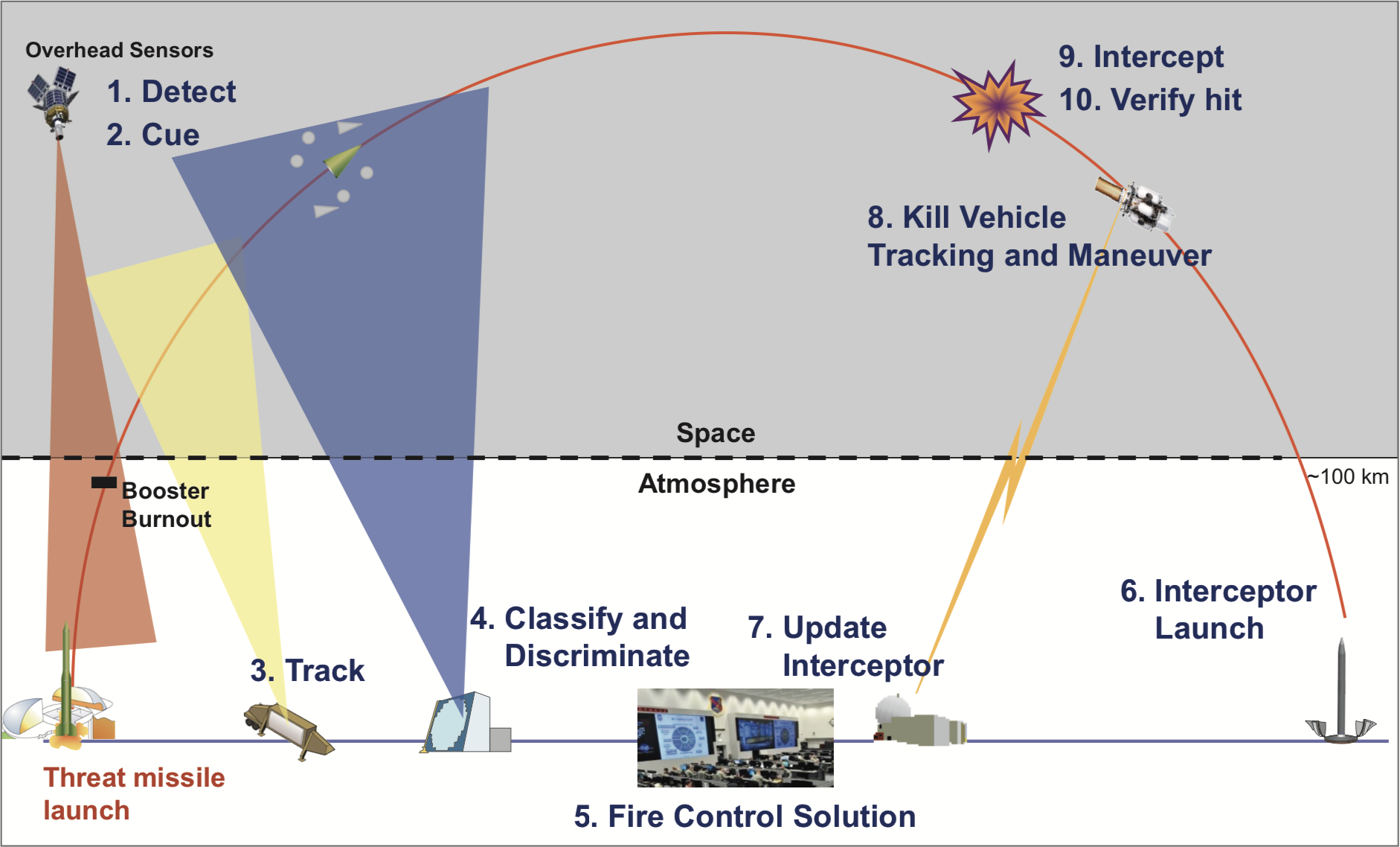}
    \caption{Ballistic Missile Acqusition and Engagement}
    \label{fig:missileInterceptDiagram}
\end{figure}

The Aegis system utilizes both local and remote subsystems (Figure \ref{fig:missileInterceptDiagram}).  The capabilities of these subsystems are realized during the acquisition, planning, tracking, and engagement of a ballistic threat.  The system utilizes external systems such as satellites, sea-based radar, ground-based tracking radar, and forward-based radar systems.  Once a threat is detected, these external subsystems transfer data to shipboard systems for threat tracking and engagement.  Local systems include radar, electro-optic/infrared  sensors, passive electronic warfare sensors, communications/link architecture, command and control systems, decoys, electronic countermeasures \cite{sysArch}.  Once the threat is tracked, it is intercepted utilizing shipboard weapons systems.  For Aegis ballistic missile defense systems, standard Missile or SM-2-Block IV missiles are utilized to engage and destroy the threat \cite{sysArch}.

Furthermore, the system communicates with various entities such as U.S. Strategic Command, U.S. Pacific Command, and fire control suites across the nation.  These systems must work in unison to achieve a common goal (threat detection and engagement).

\subsection{SoS Entites}
Four entities comprise a SoS.
\begin{enumerate}
    \item Services: Services of a SoS are the services they provide. Such services are the direct applications of the system (Ballistic missile defense).  Furthermore, the system provides indirect benefits.  These are the technological advancements made in the development and application of the SoS.  For guidance and tracking subsystems, advancements would improve a variety of civilian services.  Some of these services include the optimization of driver routes, reduction of theft and control of vehicle functions, mobility pattern recognition, improved vehicle navigation, and fleet management.
    
    \item Service Owners: Entities that have final say over service tasking.  Service owners also uphold resource allocation.  These entities are also the primary motivation in the development of the SoS.  This can be through policy decisions, technical, and performance requirements as stated in the contract.

    \item Stakeholders: SoS stakeholders are those that benefit from the capabilities of the system.  For the vehicle guidance and tracking subsystems, stakeholders are both military and civilian entities. For civilians, the technological advancements of the system have the potential of influencing the development of civilian technology.  For both civilian and defense entities, the system provides a technological advantage over adversaries and better defense potential.
    
    \item Contracts: Contracts are essential to a SoS because they establish the framework in which the SoS operates.  Contracts dictate the allocations of resources throughout the life cycle of the system.  They also organize the roles of the SoS entites.
\end{enumerate}

\subsection{Boardman and Sauser SoS Characteristics}

In order to better understand the composition of the SoS, the relationships and roles of subsystems will be investigated.

Boardman and Sauser identify a specific set of characteristics that define a SoS.  These are autonomy, belonging, connectivity, diversity, and emergence \cite{bsSoS}.

Autonomy is the capability of subsystems to complete tasks without outside control. Some systems that comprise the guidance and tracking systems are operationally independent.  An example of this is the forward-based radar system (AN/TPY-2).  The unit itself was designed as a multi-functional system.  This system can be self-cued or accept cues from outside systems (Aegis or other satellites) \cite{antpy2}.  These independent systems are networked together to create the Aegis system.

The system's contribution to national defense cements its belonging.  Guidance and tracking systems are essential when it comes to ballistic missile defense.

Connectivity measures the interoperability between the system and new systems.  The Aegis system fulfills this characteristic because of the network of interconnected systems it uses.  As mentioned previously, the system has a network of satellites for threat detection.  Upon threat detection, the data is transferred to the appropriate system for further analysis.

Boardman and Sauser describe that it is necessity for a SoS to "be incredibly diverse in its capability as a system" \cite{bsSoS}.  Ballistic missile defense systems are incredibly diverse in their capabilities.  Not only is the Aegis implemented in seagoing vessels, but land based systems as well \cite{missileDefenseAgency}.

Emergent capabilities are designed into a SoS by a virtue of factors.  These factors are defined as "preservation of constituent systems autonomy, choosing to belong, enriched connectivity, and commitment to diversity of SoS manifestations and behavior" \cite{bsSoS}  The Aegis system is continuously evolving and flaws in the system are being discovered and resolved.  The developments are essential to keep a technological advantage over adversaries.

% With a SoS, emergent behavior dare not be restricted to what can be foreseen or deliberately designed in, even if this risks greater unintended consequences, though of course these can still be tested for. A SoS must be rich in emergence because it may not be obvious what tactical functionality is required to achieve broad capability. Instead, a SoS has emergent capability designed into it by virtue of the other factors: preservation of constituent systems autonomy, choosing to belong, enriched connectivity, and commitment to diversity of SoS manifestations and behavior. The challenge for the SoS designer is to know, or learn how, as the SoS progresses through its series of stable states, to create a climate in which emergence can flourish, and an agility to quickly detect and destroy unintended behaviors, much like the human body deals with unwanted invasions.

\section{Simulation}
% Section for the simulation

The general goal behind the simulation was to represent the tracking and guidance behaviors being researched. Since there was a lot of leeway in what kind of system these behaviors could be implemented into, the decision was made to use two small-scale cars to show this behavior. The remainder of this section describes exactly how this was achieved.

\subsection{Overview}

At a large scale, this simulation would require two configurable cars and a means of controlling one of the cars. Each of the two cars would require servo motors to not only power the wheels but also set the speed of the wheels in finely quantized levels. Beyond that, each of the cars would need a way to control the servo motors and take in inputs from sensors or a controller to influence the way the servo motors are being controlled. For both cars, this would be achieved with a microcontroller, but the microcontrollers would need to fit different requirements for each of the two cars. For the car being controlled (this will be referred to as the leader) the microcontroller would need to support a wireless communication standard for communicating with the controller, such as WiFi or Bluetooth. For the car that follows the leader using sensor data (this will be referred to as the follower) the microcontroller would need to have an ADC to be able to read the sensor data. Next, both cars would need rechargeable batteries to power the servo motors and microcontrollers wirelessly. For each vehicle this battery would need to be 6V to be at a high enough voltage that the 5V regulators for the microcontrollers don't drop out, but also low enough to fit the specifications for the servo motors. The leader would need a way to emit light for the follower to track. Lastly, the follower would need sensors to sense that light and compare it to the ambient light of the environment.

\subsection{Parts Breakdown}

\begin{figure}[H]
    \centering
    \includegraphics[scale=0.05]{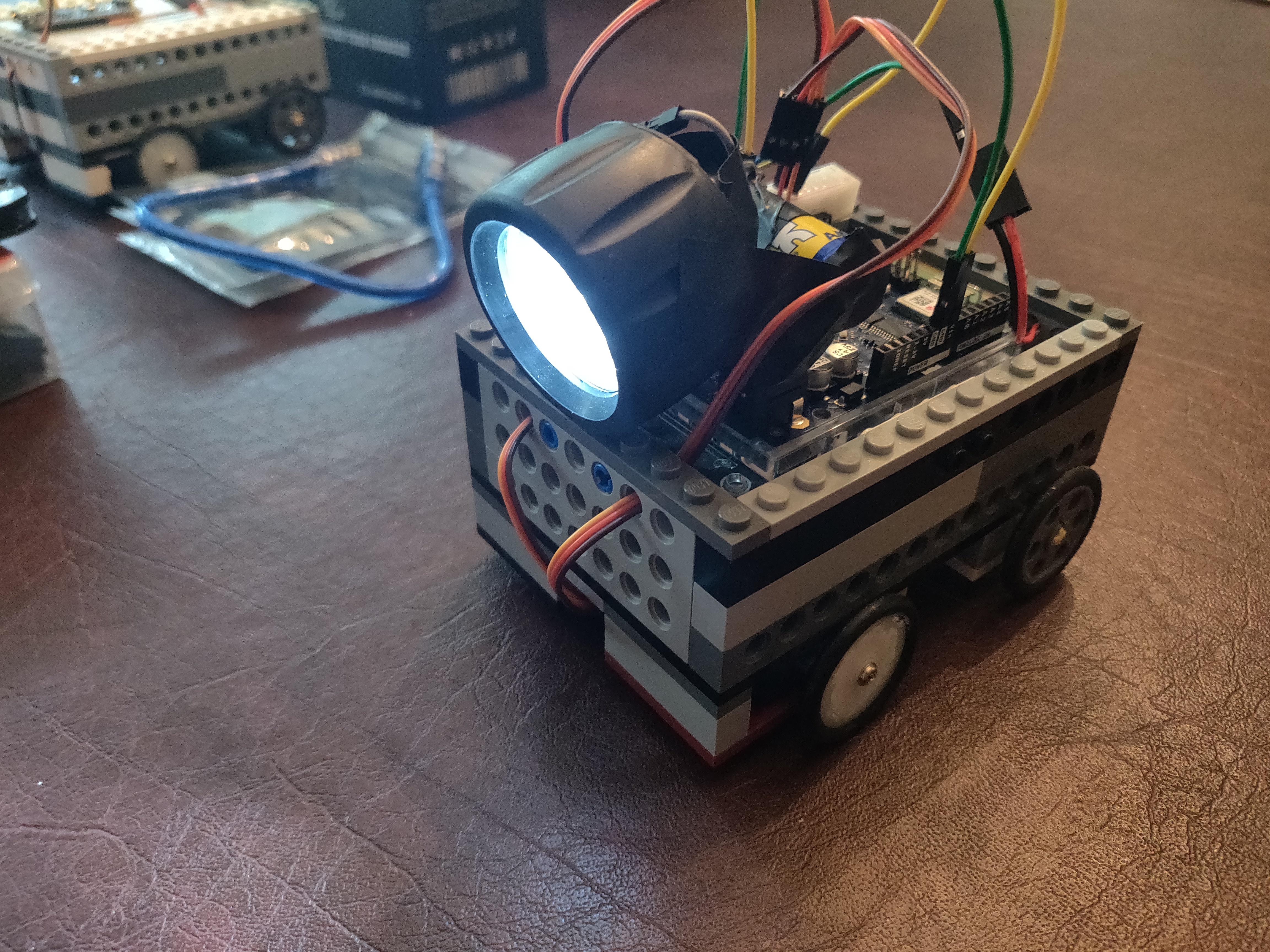}
    \caption{Leader}
    \label{fig:leader}
\end{figure}

\begin{figure}[H]
    \centering
    \includegraphics[scale=0.05]{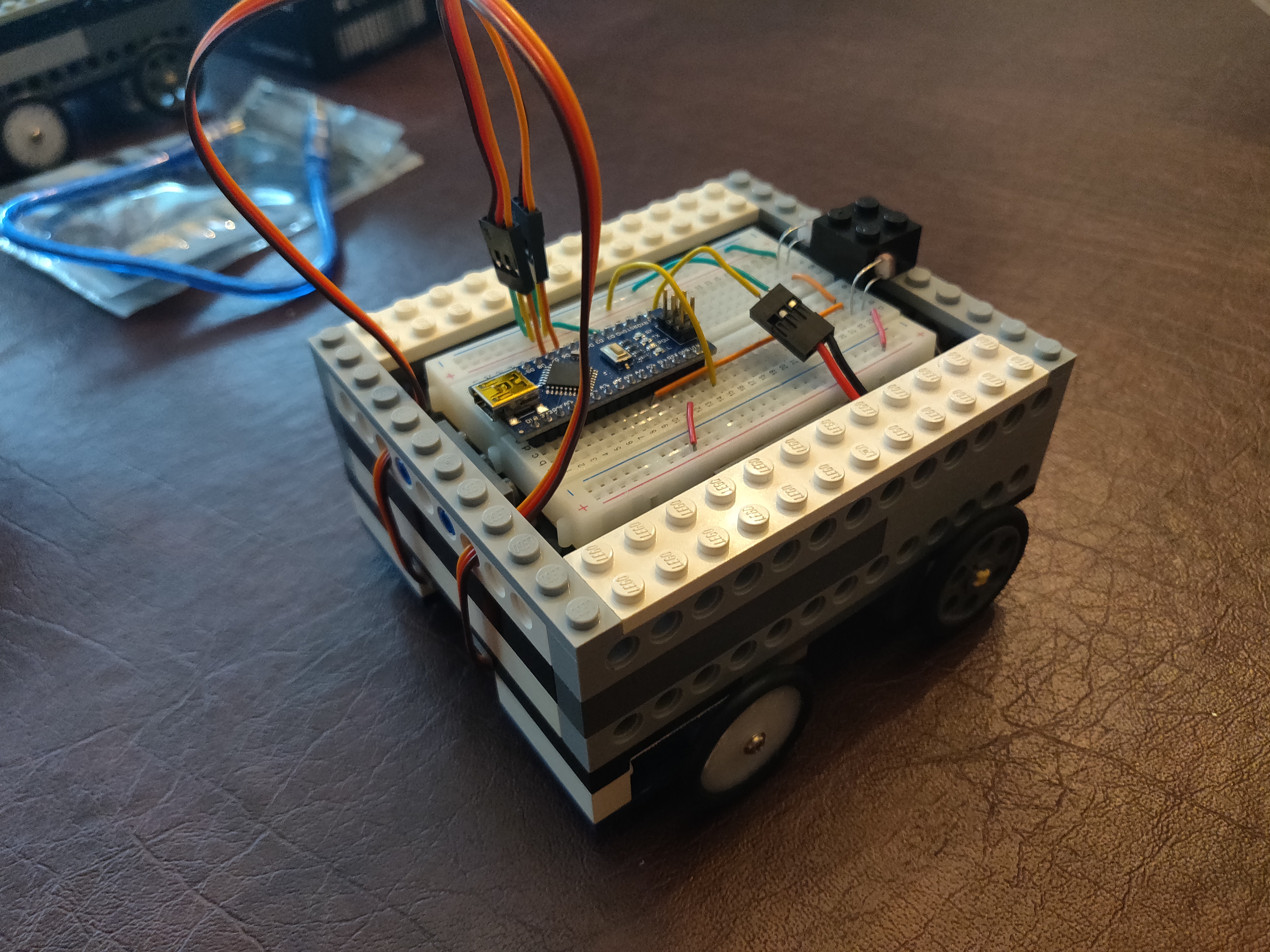}
    \caption{Follower}
    \label{fig:follower}
\end{figure}

For the servo motors, four FS90R continuous rotation servo motors were used. The primary reason for this choice was that four of these servo motors were already in the possession of one of the students. Each car would use two of these servo motors to power the rear wheels. For the leader, an Arduino Uno WiFi Rev2 was used. This microcontroller was chosen because it supports WiFi communication and it is compatible with the Arduino programming language. Since the Arduino Uno WiFi Rev2 has an ADC, this same microcontroller could have been used for the follower, however this would not have been cost effective. Instead, an Arduino Nano was used for the follower. For the batteries, standard 6V NiMH battery packs were used. These batteries were chosen because they are a standard for RC applications meaning that additional charging equipment would not be needed. For the light source, the top assembly of a generic LED flashlight was used. Since nothing unique was needed, this was a cost effective way to acquire a light source. Lastly, two photoresistors were used to compare the light levels to left and right of the follower. Since these are resistors, they can be used as a variable voltage divider. The voltage read in by the microcontroller indicates which sensor is receiving more light. In a way, this is some basic, hardware-level sensor data fusion. The leader and follower can be seen in Figures \ref{fig:leader} and \ref{fig:follower}, respectively.

\subsection{Testing Procedure}

To test the performance of the follower, various tests were performed with the leader being controlled in different ways and the two cars beginning at varying distances from each other. The overall takeaway from these tests was that the follower could follow the leader regardless of maneuvering strategies so long as the two cars began within about ten inches or less of each other. Beyond that distance, the follower had little to no ability to differentiate between the perceived light levels to left and right of the car. It is worth noting that the top speed and turning radius of the leader was artificially handicapped in software to give the follower a real chance at catching the leader. Without the turning radius limitation, the leader could likely turn all the way around and stop moving to prevent the follower from tracking it.

\section{Process Loop Mapping}
% Section on Process Loop mapping (OODA, etc.)
%  – Consider decision-making process
\begin{figure}[H]
\centering
\includegraphics[width=.8\linewidth]{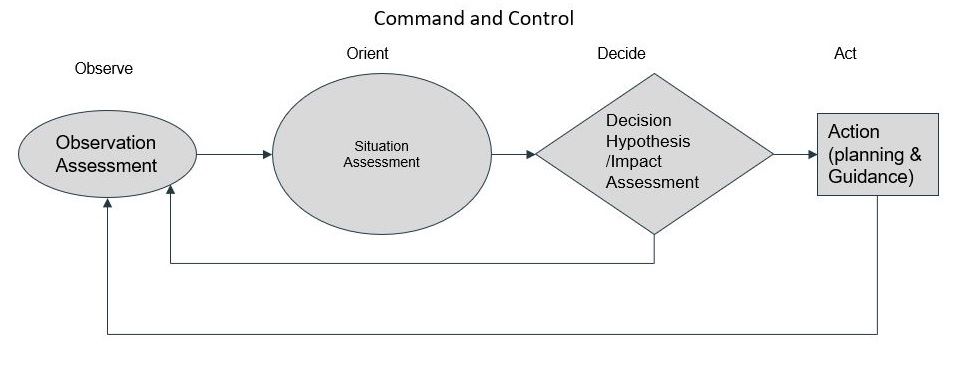}
\caption{OODA Loop}
\label{Figure1a1b}
\end{figure}

The OODA loop is a four-step approach defined by John Boyd for decision-making that focuses on filtering available information, putting it in context and quickly making the most appropriate decision while also understanding that changes can be made as more data becomes available. The four steps to an OODA loop are Observe: Understand the changing environment. Orient: Evaluate mental models, if do they still hold up under these new circumstances? If they’re still relevant,use them. If they are not, take them apart and build new ones. This is the most important step in the loop. Decide: Choose the best hypothesis from the orient phase. Act: The loop must lead to action, which can mean fully implementing the idea or just testing the hypothesis. When one loop is complete, the next loop begins. The first three steps are preparation, ensuring that the right action can be taken at the right moment.In competitive or hostile situations, decision makers should aim to complete their loop faster than the opponent. Their action resets the opponent’s loop by changing the environment and sends them back to the beginning. Decision makers need to continually challenge and adapt their mental models because the world is always changing. Models can become obsolete in a changing environment. Applying outdated models to new or unfamiliar situations can lead to disaster. To avoid disaster decision makers must continually revise mental models based on observations of the environment.

The key benefits of a OODA loop are Speed: Fighter pilots must make many decisions in fast succession. They don’t have time to list pros and cons or to consider every available avenue. Once the OODA loop becomes part of their mental toolboxes, they should be able to cycle through it in a matter of seconds. Speed is a crucial element of military decision making. Using the OODA loop in everyday life, we probably have a little more time than a fighter pilot would. But Boyd emphasized the value of being decisive, taking initiative, and staying autonomous. These are universal assets and apply to many situations. Comfort With Uncertainty: Uncertainty does not always equate to risk. A fighter pilot is in a precarious situation, where there will be gaps in their knowledge. They cannot read the mind of the opponent and might have incomplete information about the weather conditions and surrounding environment. They can, however, take into account key factors such as the opponent’s nationality, the type of airplane they are flying, and what their maneuvers reveal about their intentions and level of training. If the opponent uses an unexpected strategy, is equipped with a new type of weapon or airplane, or behaves in an irrational, ideologically motivated way, the pilot must accept the accompanying uncertainty. However, Boyd belabored the point that uncertainty is irrelevant if we have the right filters in place. Unpredictability: Using the OODA loop should enable us to act faster than an opponent, thereby seeming unpredictable. While they are still deciding what to do, we have already acted. This resets their own loop, moving them back to the observation stage. Keep doing this, and they are either rendered immobile or forced to act without making a considered decision. So, they start making mistakes, which can be exploited. Testing: A notable omission in Boyd’s work is any sort of specific instructions for how to act or which decisions to make. This is presumably due to his respect for testing. He believed that ideas should be tested and then, if necessary, discarded.

\section{Conclusion}

 Construction and simulation of working model is done for demonstrating the Vehicle tracking and Guidance system. Project performance illustrates on C2 aspects like Command, Control, Communication, and Sensor Data Fusion. Working model performed well and the system can be improved by using latest Communication technologies, using various types of high precision sensors and high speed processors.

\end{document}